# AI in Archival Science - A Systematic Review

Gaurav Shinde
Department of Engineering
San José State University
San Jose, CA, USA
gauravyeshwant.shinde@sjsu.edu

Tiana Kirstein
School of Information
The University of British Columbia
Vancouver, BC, Canada
tdk846@student.ubc.ca

Souvick Ghosh
School of Information
San José State University
San Jose, CA, USA
souvick.ghosh@sjsu.edu

Patricia C. Franks
School of Information
San José State University
San Jose, CA, USA
patricia.franks@sjsu.edu

## Abstract

The rapid expansion of records creates significant challenges in management, including retention and disposition, appraisal, and organization. Our study underscores the benefits of integrating artificial intelligence (AI) within the broad realm of archival science. In this work, we start by performing a thorough analysis to understand the current use of AI in this area and identify the techniques employed to address challenges. Subsequently, we document the results of our review according to specific criteria. Our findings highlight key AI driven strategies that promise to streamline record-keeping processes and enhance data retrieval efficiency. We also demonstrate our review process to ensure transparency regarding our methodology. Furthermore, this review not only outlines the current state of AI in archival science and records management but also lays the groundwork for integrating new techniques to transform archival practices. Our research emphasizes the necessity for enhanced collaboration between the disciplines of artificial intelligence and archival science.

# 1. Introduction

## 1.1. Artificial Intelligence

The landscape of Artificial Intelligence (AI) has witnessed notable transformations in both its conceptualization and practical applications in recent years. Originating in the mid-20th century with a focus on rule-based systems emulating human intelligence, the contemporary definition of AI has undergone a paradigm shift. In 2007, Stanford professor John McCarthy succinctly defined AI as "the science and engineering of making intelligent machines." Fast forward to 2021, the goal of AI has developed into a machine that can think like humans and mimic human behaviors, including perceiving, reasoning, learning, planning, predicting, and so on (Xu et al., 2021). The current emphasis is on machines' capability to learn and adapt from data, a transformation largely attributed to the ascendancy of machine learning methodologies. In recent years, deep learning, a subset of machine learning, has gained prominence, empowering computers to process and understand vast amounts of complex data. The development of neural networks, particularly those with multiple layers, has revolutionized AI capabilities, enabling significant strides in tasks like image recognition, natural language processing (NLP), and strategic game theory. Figure 1 highlights the influential work in AI over the years. Furthermore, the integration of AI across diverse sectors, ranging from healthcare to finance and autonomous vehicles, highlights the practicality and versatility of these intelligent systems in addressing real-world challenges. However, as AI continues to advance, concerns surrounding ethical considerations, transparency, and the broader societal implications of these technologies persist, adding depth to the ongoing discourse within the field.

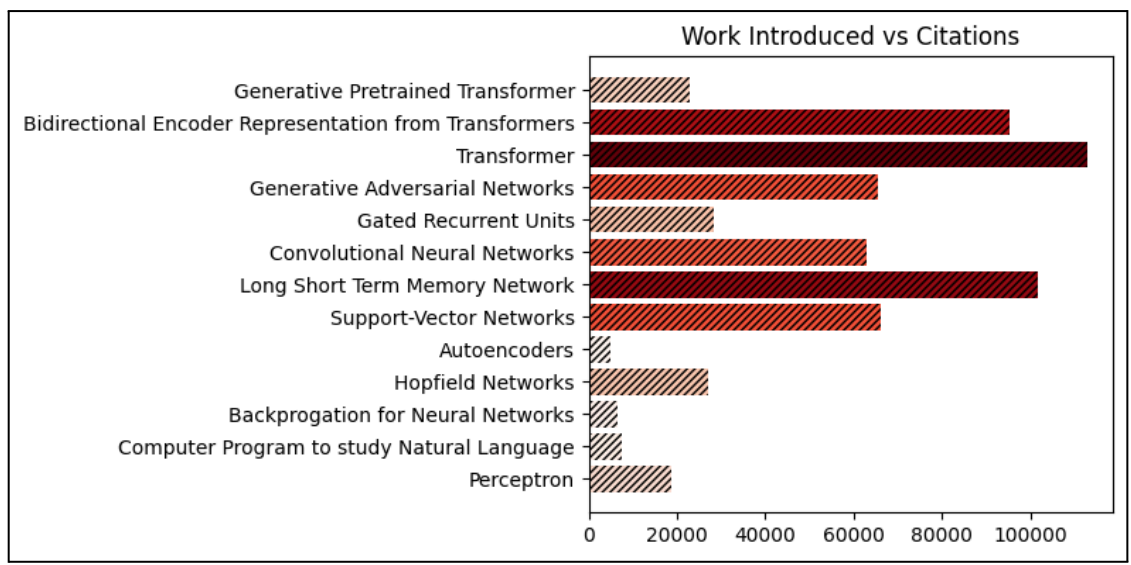

**Fig. 1** Influential work in the field of Artificial Intelligence from 1958 to 2020

## 1.2. Records Management and Archival Science

Records management is the "systematic design, implementation, maintenance and administrative control of a framework for the making and keeping of records to ensure efficiency and economy in their creation, use, handling, control, maintenance, and disposition" (InterPARES Trust, 2023). Its purpose is to capture and maintain evidence of business and organizational functions. Archival science is the "systematic body of knowledge that supports the practice of appraising, acquiring, authenticating, preserving, and providing access to recorded materials" (InterPARES Trust, 2023). It provides the foundational theory behind the work that archivists, records managers, and other information professionals do to preserve the documentary evidence of people, institutions, organizations, businesses, and communities that have long-term value. Archives and other records repositories support the construction of social memory, ensure the accountability of government and non-government institutions by preserving records and making them available to the citizens, shareholders, boards, and other constituents as is legally and ethically appropriate, serve as memory institutions for communities, and support other scholarly, administrative, and personal research (Gilliland-Swetland, 2000). Together, records management and archival science ensure access to information from its creation to its designated end, whether that be destruction or perpetual preservation.

## 1.3. Benefits of integrating AI and Archives

Integrating AI with archive science transforms how collections are handled, accessed, and preserved, providing several advantages that broaden the field's capabilities and efficiency. AI-powered techniques improve records' accessibility by automating document transcription, making them searchable for a wider audience. They boost discoverability by using powerful algorithms that identify patterns. AI also plays an important role in indexing and processing large volumes of archival material (Carter et al., 2022) and historical document analysis (Lombardi & Marinai (2020). It also solves the issues of managing a large amount of digital information by automated categorization and metadata assignment. This integration not only reduces costs by automating mundane processes, but it also creates new opportunities for academic study by revealing hidden trends and patterns through data analysis. Furthermore, AI's language processing skills remove linguistic boundaries, making archives accessible to a worldwide audience and promoting global study.

## 1.4. Research Questions

- ***RQ1:*** What AI techniques have been used to execute archives and records management functions?
- ***RQ2:*** What archives and records management functions have been executed by AI?
- ***RQ3:*** How could advancements in AI potentially transform archival and records management practices?

The rest of the paper is organized as follows: Section 2 introduces the problem statement and Section 3 explains the systematic review approach. Sections 4 and 5 look into AI technologies and the applications of AI in records management and archives, respectively. Section 6 answers the research questions and Section 7 concludes the paper and highlights future research directions.

# 2. Problem Statement

As we have moved into an increasingly digital society, the number of digital records being produced has grown exponentially. This growth has presented challenges in nearly all aspects of records management and archival work, from retention and disposition, to appraisal, arrangement and description, and access and preservation. These records are adding to the pre-existing backlog of paper records, for which there is an increasing demand for digitization and online access. The vulnerability of digital records demands that they be processed and preserved upon acquisition by archives due to quick-deteriorating magnetic and optical media storage and their dependency on specific software (Matlala, 2019). One of the primary types of digital records that require specialized attention is e-mail correspondence. Though most e-mail systems use automation to sort correspondence, refiling correspondence by humans is often necessary to meet the requirements of organizations' records management schemas (Lappin, 2020). However, this refiling is often infrequent and haphazard and rarely meets retention and disposition requirements. On the archival side, accessioning and providing access to e-mails is difficult due to the large volume of correspondence and the need to screen it for sensitive, confidential, or legally restricted information (Schneider et al., 2019). Yet there is a growing demand from researchers for access to both born-digital and digitized records, regardless of privacy and security concerns (Jaillant, 2022a). In addition to solving pre-existing issues, AI provides an opportunity to access records in new ways. AI-powered technologies like Handwritten Text Recognition (HTR), Optical Character Recognition (OCR), Facial Recognition, Language Translation and other similar systems allow users to perform various kinds of activities.

## 3. Systematic Review Approach

A systematic review is a meticulous approach to examining scientific publications. It involves conducting a thorough search for relevant research, assessing their quality, and synthesizing the findings of these studies using predetermined criteria. This technique is intended to reduce bias while providing a trustworthy summary of the evidence on a specific research subject. Systematic reviews are significant because they collect data from numerous sources, providing a more comprehensive knowledge of study findings than individual studies. By critically appraising and synthesizing existing research, systematic reviews can provide clarity on conflicting results and guide practitioners in making evidence-based decisions. Furthermore, they play an important role in identifying research gaps and creating goals for future studies.Our systematic review is segmented into three principal modules: Planning, Preparation, and Analysis. Figure 2 depicts the steps involved in our systematic review process. Each module serves a distinct purpose and will be described in detail below.

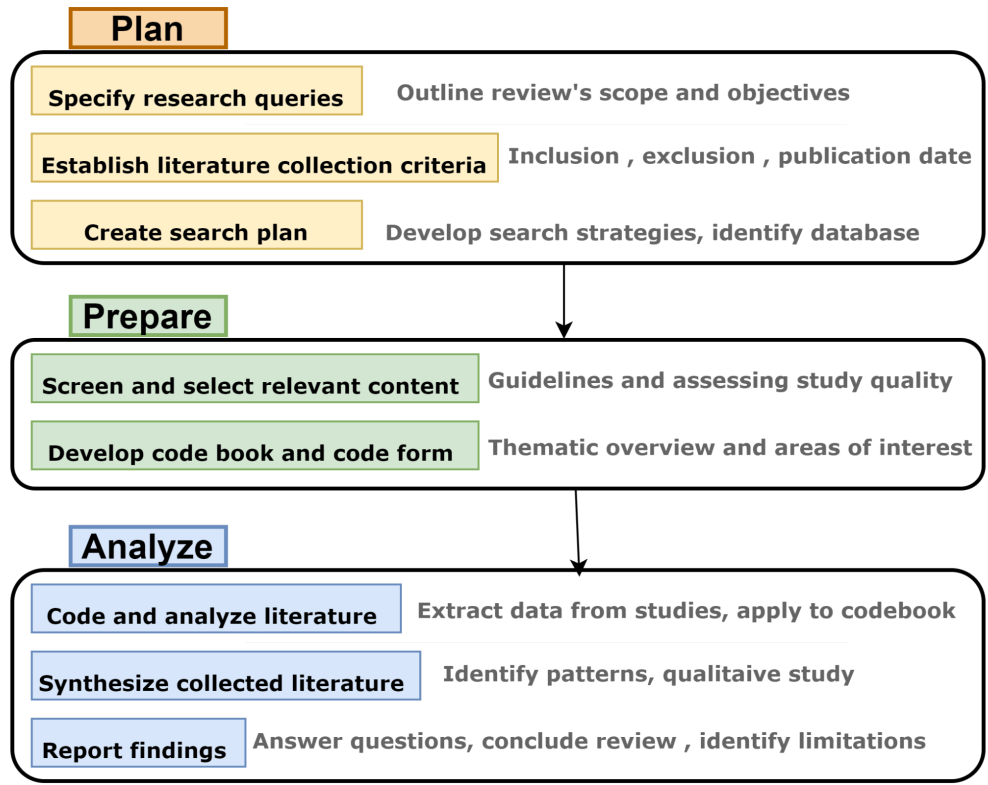

**Fig. 2** Steps involved in the Systematic Review

## 3.1. Planning

The planning stage commenced with a thorough ideation session on the optimal strategy for carrying out a survey in this field. This led to the creation of a document where the authors meticulously collated pertinent studies from several databases. Each research paper was assessed using the below criteria:

1. The purpose of the research
2. The overall idea of the paper
3. The problem addressed in the study
4. The methodological approach followed in the study
5. Findings of the study
6. Originality of the study
7. Social implications of the work
8. Future directions

These criteria helped us identify patterns and trends that were closely followed in most papers. As we wanted to discover state-of-the-art research happening at the intersection of Archival Science and Artificial Intelligence, we analyzed publications from the past five years (2019-2023). We exclusively sourced research papers written in English from all the databases in our search. The databases we utilized mainly are ACM Digital Library, IEEE Xplore, Internet Archive, Elsevier, SpringerLink, ScienceDirect, Scopus, Web of Science, and PubMed Central. Search terms used included "Archives and Artificial Intelligence," "Records Management and Artificial Intelligence," "Archives and Machine Learning,"

"Records Management and Machine Learning," "Digital Archives," "Neural Networks applications for Archival Science," AI Archives" and "AI ML Records." We began our review with more than 2000 articles.. Our initial screening, involving filtering by year, resulted in around 1,000 papers. Next, we conducted more rounds of screening to determine relevance. Studies that only briefly mentioned AI as a possible solution to minor issues were excluded. Two more papers were added, through citation chaining, which were relevant to our topic of review but were published before 2019. The final list of papers contained 45 papers, where the papers chosen were those that used AI or described a way in which AI could be applied to solve a specific issue or increase access to archives. Figure 3 illustrates our selection criteria.

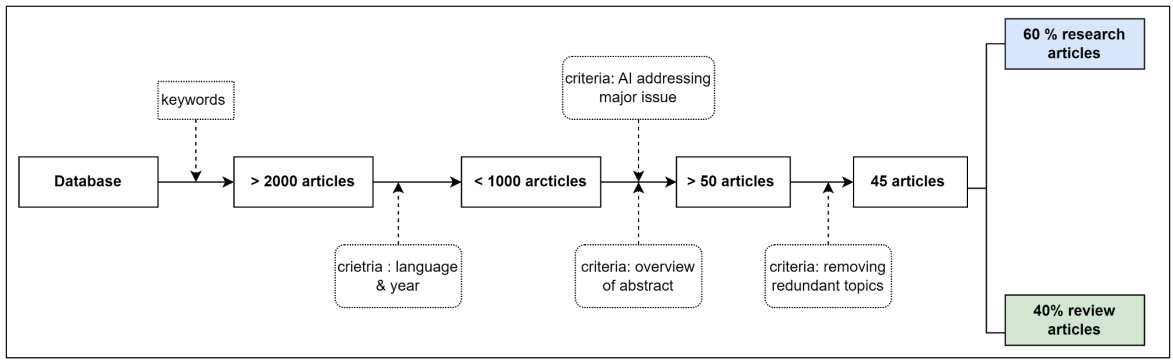

**Fig. 3** Paper Selection Process

## 3.2. Preparation

During the preparation phase, our focus was to pinpoint themes and subthemes pertinent to both Artificial Intelligence and Archival Science. For AI, we have elaborated on these themes and subthemes in Section 3; for Archival Science (including records management), these are detailed in Section 5.

## 3.3. Analyze

The major focus in this phase was to uncover patterns and trends which have been highlighted below.

### 3.3.1. Publication Characteristics

Figure 4 displays the number of articles reviewed from various database sources, and Figure 5 indicates the annual count of conference and journal articles reviewed from 2019 to 2023.

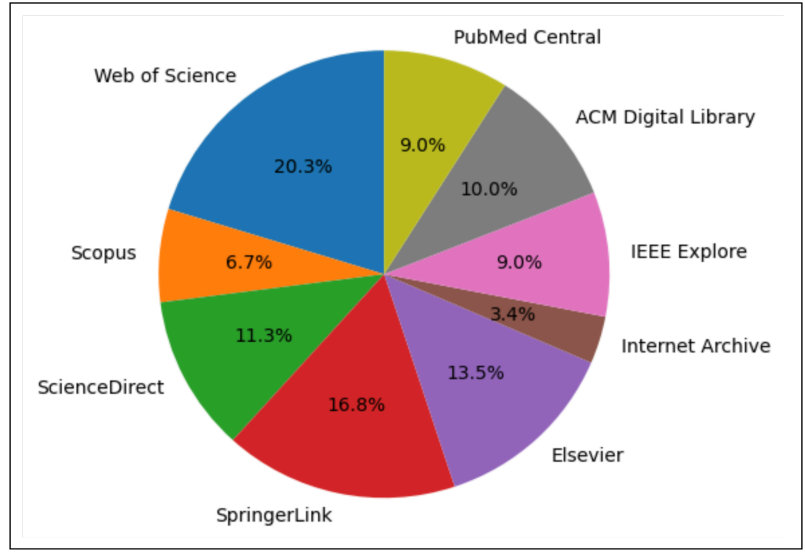

**Fig. 4** Distribution of Papers Reviewed Across Different Databases

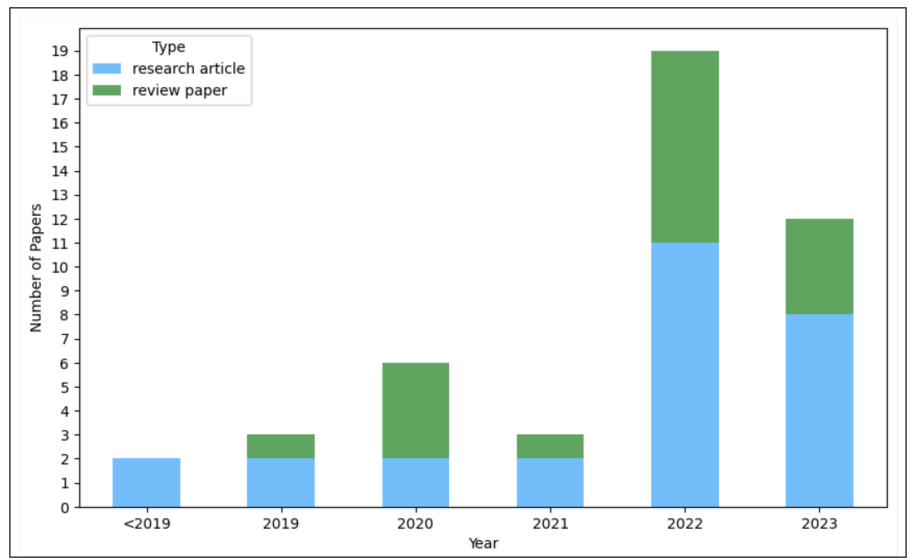

**Fig. 5** Distribution of research articles vs review papers reviewed for each year

# 4. Artificial Intelligence Approaches

The integration of Artificial Intelligence (AI) into the domain of archives and records management represents a paradigm shift in how we approach information curation, preservation, and accessibility. This section situates its inquiry within the burgeoning field of AI-enhanced archival practices, reviewing key contributions and pioneering efforts. It describes the strides made in incorporating machine learning, natural language processing, and other AI technologies to enhance archival systems. This section sets the stage for a detailed exploration of the transformative impact of AI on archival methodologies, highlighting the interplay between technological advancements and the evolving nature of records management. Table 1 organizes our literature review into distinct AI subdomains, enabling readers to navigate efficiently. Figure 6 presents the number of papers reviewed per field in machine learning, as well as the number reviewed for each technique within those fields.

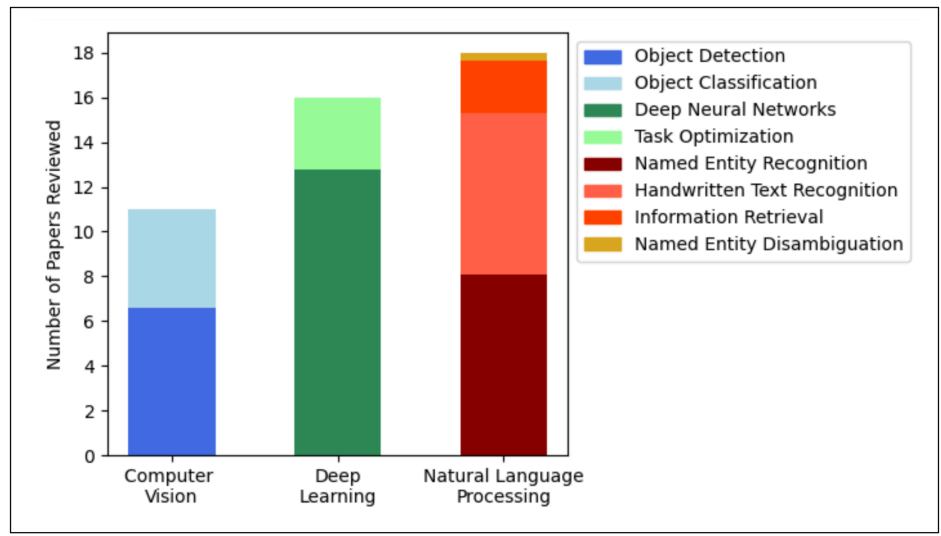

**Fig. 6** Distribution of Themes and Subthemes for AI Techniques

## 4.1. Natural Language Processing

Natural Language Processing (NLP) is a field of computer science and linguistics focused on the interaction between computers and human language. It involves programming computers to process and analyze large amounts of natural language data. This section examines the application of NLP in archival studies, detailing how NLP techniques are utilized to enhance archival research and management. It categorizes the literature on NLP in archives into specific areas: Named Entity Recognition (NER), Handwritten Text Recognition (HTR), Named Entity Disambiguation (NED), and Information Retrieval (IR), providing a structured overview of how these technologies are employed in the field.

### 4.1.1. Named Entity Recognition

Named Entity Recognition (NER) is a subfield of Natural Language Processing (NLP) that involves identifying and classifying key information in text into predefined categories. These categories typically include names of persons, organizations, locations, expressions of times, quantities, monetary values, and more. NER is essential for understanding the context of documents by extracting meaningful and structured information from unstructured text, making it a valuable tool for data analysis, information retrieval, and content organization. This process enables more efficient handling and organization of large volumes of text, particularly in areas like archives and records management.

| Themes | Subthemes | Citations |
| --- | --- | --- |
| Natural Language Processing | Named Entity Recognition | Haffenden et al., 2023; Carter et al., 2022; Fan et al., 2022; Alothman & Wahab Sait, 2022; Schneider et al., 2019; Modiba, 2023; Ehrmann et al., 2023 |
| | Handwritten Text Recognition | Terras, 2022; Nockels et al., 2022; Rolan et al., 2019; Lappin, 2020; Kumar et al., 2021; Yang, 2023; Bazán-Gil, 2023 |
| | Named Entity Disambiguation | Haffenden et al., 2023 |
| | Information Retrieval | Alothman & Wahab Sait, 2022; Modiba, 2023 |
| Traditional Machine Learning | Machine Learning | Bottrighi et al., 2022; Michaud et al., 2023; Liu et al., 2023; Triantafyllou et al., 2023; Lafia et al., 2021; Cheng et al., 2015; Shang et al., 2019 |
| Deep Learning | Deep Neural Networks and Task optimization | Suissa, Elmalech, & Zhitomirsky-Geffet, 2022; Rolan et al., 2019; Lombardi & Marinai, 2020; Ehrmann et al., 2023; Deckers & Potthast, 2022; Yang, 2023; Adewoye et al., 2020; Randby & Marciano, 2020 |
| Computer Vision | Object Detection and Classification | Carter et al., 2022; Männistö et al., 2022; Yang, 2023; Adewoye et al., 2020; Bazán-Gil, 2023; Okamoto et al., 2023; Cheng et al., 2015; Randby & Marciano, 2020 |
| Explainable Artificial Intelligence | | Jaillant, 2022b; Bottrighi et al., 2022; Davet, Hamidzadeh, & Franks, 2023 |

**Table 1**. Artificial Intelligence application avenues

Haffenden et al. (2023) introduced KB-BERT, a BERT (Bidirectional Encoder Representations from Transformers) which is fine-tuned for NER tasks. This model is also utilized for text classification merging BERT's capabilities with bootstrapping techniques, and enhanced Optical Character Recognition (OCR) cohesion through the use of Next Sentence Prediction (NSP). Carter et al. (2022) outline a 2-phase supervised machine learning algorithm that leverages tools like Google AutoML Vision. This algorithm is applied to process Table of Contents (TOC) pages, identifying subject index boxes, volume page references, and dates. The end result is an application that functions as a comprehensive CRUD (Create, Read, Update, Delete) interface. Fan et al. (2022) made use of a technique known as Data-driven and Circulating Archival Processing (DCAP). This paper also involves Bayesian models, which play a crucial role in document categorization. Alothman & Wahab

Sait (2022) focus on three primary areas: building a data extraction model, developing a NER classifier, and implementing a ranking algorithm for document retrieval. For the NER classification, the study utilized Multinomial Naive Bayes. To rank documents, it integrated a combination of algorithms, including PageRank, Hyperlink-Induced Topic Search (HITS), and Stochastic Approach for Link-Structure Analysis (SALSA). Schneider et al. (2019), through extensive study, demonstrated how Email Preservation and Disclosure Platform (ePADD) can significantly help with automated entity recognition and search functionalities.

### 4.1.2. Handwritten Text Recognition

Handwritten Text Recognition (HTR) refers to the ability of a computer system to interpret and understand human handwriting. It involves converting handwritten characters into a digital text format that computers can process and analyze. This technology has made significant strides due to advances in artificial intelligence and machine learning, resulting in improved accuracy when recognizing a wide array of handwriting styles. The application of HTR spans various domains, notably in archives and records management. Terras (2022) introduced the Transkribus platform, which employs HTR to decipher words within segmented text lines, even when overlapping probabilities are present. Furthermore, Nockels et al. (2022) delved into the application of HTR, placing particular emphasis on the Transkribus platform. Lappin (2020) explored the use of AI in determining the importance of emails, shedding light on their relevance in contemporary communication dynamics. Kumar et al. (2021) proposed a text-based image retrieval system designed to enhance the retrieval of pertinent images from a diverse range of text documents. Lastly, Yang (2023) recently introduced an enhanced text-to-video retrieval model tailored for audiovisual archives, demonstrating the continued evolution and relevance of HTR in multimedia information management.

### 4.1.3. Named Entity Disambiguation and Information Retrieval

Named Entity Disambiguation (NED) is the process of resolving references to ambiguously named entities, such as people, places, or organizations, to their specific and correct identities. In archives and records management, NED helps ensure accurate indexing and retrieval of information by distinguishing between different entities with similar names, enhancing data organization and search capabilities. On the other hand, Information retrieval (IR) is the process of searching for and obtaining relevant information from a repository or database. In archival studies, it is employed to efficiently locate, retrieve, and access specific records or documents, ensuring timely and accurate access to historical data. Alothman & Wahab Sait (2022) implemented a ranking algorithm for efficient document retrieval. Modiba (2023) discusses the several advantages of efficient information retrieval with a specific emphasis on records management.

## 4.2. Traditional Machine Learning

Traditional machine learning is a subset of AI that involves the use of statistical techniques to enable computers to learn from and make predictions based on data. Unlike deep learning, traditional ML typically relies on hand-engineered features and simpler algorithms. This section explores the integration of traditional ML in the realm of archives.

### 4.2.1. Machine Learning

Machine learning (ML) enables computers to learn from data without explicit programming. It is notable for its ability to adapt and improve over time. This technology is increasingly pivotal in advancing various research fields due to its efficiency. It offers several use cases in archival studies. Bottrighi et al. (2022) made use of various models such as Logistic Regression, Support Vector Machine, Decision Tree, and Random Forest to find patterns in patient profiles. Michaud et al. (2023) proposed a labeling function method that involves segmentation, feature computation, and classification to reduce noise in bird song data. Liu et al. (2023) deduced crucial wind information using eight different ML models. Lafia et al. (2021) developed a computational model to identify and characterize a broad range of data curation activities to explore the impact of organizational changes on curation practices. Shang et al. (2019) introduced an efficient method to classify archives using XGBoost and Spark computing.

## 4.3. Deep Learning

Deep Learning involves algorithms inspired by the structure and function of the brain to model complex patterns and make decisions with high accuracy. In the field of archives and records management, it has transformed how vast amounts of data are processed and retrieved. Deep learning is considered state-of-the-art for modern archival processes.

### 4.3.1. Deep Neural Networks

Deep Neural Networks (DNNs) are a complex architecture modeled after the human brain's network of neurons. They consist of multiple layers of interconnected nodes, each capable of performing specific computations. This makes them highly effective for applications requiring nuanced understanding. Rolan et al. (2019) explore various deep learning techniques in the context of recordkeeping studies. Lombardi & Marinai (2020) present an overview of how neural nets are used in historical document analysis. This serves as a comprehensive reference for researchers and practitioners working at the intersection of deep learning and historical studies. Ehrmann et al. (2023) offer insights into the strengths and weaknesses of deep architectures, providing recommendations for optimal pre-processing and transferability. Deckers & Potthast (2022) introduce WARC-DL to efficiently process large web archives, thereby providing a scalable solution for the training and inference of DNNs using web archive data. Adewoye & Wahab Sait (2020) implemented DAIRE (Deep Archival Image Retrieval Engine), which is an image exploration tool based on neural networks designed to enable people to search web archives using visual content instead of traditional text-based queries.

## 4.4. Computer Vision

Computer vision enables machines to interpret and process visual data from the world, akin to human vision. In archives and records management, computer vision plays a crucial role by automating the digitization and organization of archival materials, enhancing searchability, and aiding in the preservation and analysis of historical documents and images.

### 4.4.1. Object Detection

In the context of computer vision, object detection refers to the technology that locates objects within digital images or videos, distinguishing and classifying them from their background. Carter et al. (2022) utilized Google AutoML vision to process Table of Content pages. Männistö et al. (2022) introduced the Automatic Image Content Extraction (AICE) framework. This framework accomplishes various tasks such as salience estimation, scene and event recognition, and pose estimation. Yang (2023) applied different Transformer architectures like MMT and MFT to audio-visual archives. Bazán-Gil (2023) discusses the implementation of object detection models for tasks like content annotation and facial recognition. Okamoto et al. (2023) constructed Image-Text pairs dataset from books using Object Detection, Optical Character Recognition, and Layout Analysis.

## 4.5. Explainable Artificial Intelligence (XAI)

Explainable AI (XAI) refers to AI systems designed to provide human-understandable explanations of their operations and decisions. In archival studies, XAI plays a pivotal role by offering transparency and insights into AI-driven processes used for sorting, categorizing, and interpreting archival materials. This not only enhances trust in automated archival systems but also aids archivists in understanding the rationale behind AI-generated predictions. Jaillant (2022b) and Bottrighi et al. (2022) highlight the importance of white box models over black box models and also provide practical implications. Davet, Hamidzadeh, & Franks (2023) discuss the common goals between XAI and Paradata.

# 5. Archival Application Areas - Role of AI

The literature on the applications of AI in records management and archives can be broadly divided into four themes: Records Management, Archival Processing, Access and Use, and Professional Perspectives. Table 2 divides the literature amongst these themes. Each theme is broken into two to four subthemes, which are expanded upon below.

| Themes | Subthemes | Citations |
|---|---|---|
| Records Management | Classification | Cheng et al., 2015; Ehrmann et al., 2023; Haffenden et al., 2023; Jaillant, 2022a; Lappin, 2020; Männistö et al., 2022; Michaud, Le Cesne, & Haupert, 2023; Modiba, 2022; Rolan et al., 2019; Shang et al., 2019; Triantafyllou et al., 2023 |
| | Data Collection and Management | Davet, Hamidzadeh, & Franks, 2023; Jo & Gebru, 2020; Lafia et al., 2021; Oliver et al., 2023; Roberts & Montoya, 2022 |
| | Retention and Disposition | Modiba, 2022; Rolan et al., 2019 |
| | General Records Management | Alothman & Wahab Sait, 2022; Modiba, 2023; Xie, Siyi, & Han, 2022 |

| Archival Processing | Appraisal | Colavizza et al., 2021; Deckers & Potthast, 2022; Fan et al., 2022; Lappin, 2020; Schneider et al., 2019 |
|---|---|---|
| | Arrangement and Description | Bazán-Gil, 2023; Colavizza et al., 2021; Davet, Hamidzadeh, & Franks, 2023; Haffenden et al., 2023; Jaillant, 2022b; Lombardi & Marinai, 2020; Randby & Marciano, 2020 |
| | Preservation | Cushing & Osti, 2022; Kusumawati & Salim, 2022; Modiba, 2022; Taurino & Smith, 2022 |
| Access and Use | Privacy | Cushing & Osti, 2022; Jaillant, 2022a; Jaillant, 2022b; Jaillant & Rees, 2022; Jo & Gebru, 2020; Li & Fleischmann, 2020; Modiba, 2023; Roberts & Montoya, 2022; Schneider et al., 2019 |
| | Records Retrieval | Adewoye et al., 2020; Alothman & Wahab Sait, 2022; Kumar et al., 2021; Bazán-Gil, 2023; Carter et al., 2022; Cheng et al., 2015; Chrons & Sundell, 2011; Colavizza et al., 2021; Ehrmann et al., 2023; Fan et al., 2022; Haffenden et al., 2023; Männistö et al., 2022; Nockels et al., 2022; Randby & Marciano, 2020; Suissa, Elmalech, & Zhitomisky-Geffet, 2021; Terras, 2022; Yang, 2023 |
| | Use of Records in Research | Bottrighi et al., 2022; Jaillant, 2022b; Liu et al., 2023; Michaud, Le Cesne, & Haupert, 2023; Okamoto et al., 2023; Suissa, Elmalech, & Zhitomisky-Geffet, 2021 |
| | Public Access | Branting et al., 2023; Baron, Sayed, & Oard, 2022 |
| Professional Perspectives | Archival Education | Li & Fleischmann, 2020; Poole & Diaz, 2022; Suissa, Elmalech, & Zhitomisky-Geffet, 2021 |
| | User Perceptions | Cushing & Osti, 2022; Modiba, 2023; Xie, Siyi, & Han, 2022 |

**Table 2**. Archival application areas

## 5.1. Records Management

The literature totaling twenty-two papers relating to records management, as in the systematic and administrative control of records before they are dispositioned, can be divided into four subthemes. Subtheme one is dedicated to the classification of records, which is the organization of materials into categories according to a scheme. Subtheme two discusses data collection and management, which refers to the acquisition, control, protection, and delivery of data. The third subtheme addresses retention and disposition, which is the maintenance of records for a period of time and their subsequent destruction or transfer. The final subtheme is general records management, which contains literature that refers to a specific records management function.

### 5.1.1. Classification

Twelve papers presented user AI in relation to the classification of records. Four papers present AI and ML models that can be used for classification. Cheng et al. (2015) introduce a method that accurately recognizes and classifies scenes within videos in audiovisual archives, enabling better retrieval. Similarly, Männistö et al. (2022) introduce Automatic Image Content Extraction (AICE), which can be used to analyze, classify, and search large image archives. Shang et al. (2019) present an ML method for classifying textual records, and Haffenden et al. (2023) introduce an NLP model used in the National Library of Sweden that can be used for the classification of digital materials. Ehrmann et al. (2023) provide a literature review of approaches to NER in digitized records for the purpose of identifying and classifying people, organizations, and locations of interest. Four papers present either one or more case studies in which AI is effectively applied for classification. Lappin (2020) examines numerous projects that are using ML to classify and appraise emails. The research of Rolan et al. (2019) also addresses emails, providing case studies of AI projects that address the classification and de-duplication of emails. Modiba's (2022) case study investigates the useability of AI to improve records management functions at the Council for Scientific and Industrial Research (CSIR) in South Africa and finds that automated classification made records management functions more efficient. Triantafyllou et al. (2023) present a case study in which ML is used to classify records according to subjects at the University of West Attica. Two papers present the perspectives of those using records for research. Michaud et al. (2023) use ML to reduce noise and improve classification in a bird song database, while Jaillant (2022a) investigates the barriers to the access of born-digital and digitized records, proposing AI classification of records as a solution.

### 5.1.2. Data Collection and Management

Five papers fall under the subtheme of Data Collection and Management. Lafia et al. (2021) present a schema and computational model that identifies and characterizes a broad range of data curation activities to explore the impact of organizational changes on curation work practices. Jo & Gebru (2020) argue that data collection for ML could adopt methodologies from archives and libraries to address current issues surrounding consent, power, inclusivity, transparency, ethics, and privacy. In a similar vein, Davet et al. (2023) argue that paradata should be used to explain how AI is being used in archives to ensure that archival ethic principles are maintained. Two papers also address Indigenous data sovereignty. Oliver et al. (2023) use a literature review to explore the topic of data cultures, which includes data-related skills and attitudes, data sharing, data use/reuse, and data ethics and governance, with a specific focus on Indigenous perspectives. Roberts & Montoya (2022) explore how data sharing and data mining affect indigenous cultures, with a focus on data access, privacy rights, and data governance.

5.1.3. Retention and Disposition

Only two papers addressed the applications of AI for the retention and disposition of records. Modiba (2022), in researching the usability of AI to improve records management in the CSIR, found that AI was able to identify records for retention and disposition according to retention schedules. Rolan et al. (2019) presented case studies in which AI was used to de-duplicate emails, classify records for retention and disposition, and complete other records management tasks.

5.1.4. General Records Management

These three papers explore the uses of AI in records management but do not address a specific records management activity. Two of the papers focus specifically on the perspectives of records management professionals, with Modiba (2023) investigating user perceptions on the potential uses of AI and Xie et al. (2021) studying how records management professional competencies align with the development of AI. Meanwhile, Alothman & Wahab Sait (2022) provide a framework for using NLP techniques to manage Arabic and English bilingual documents.

## 5.2. Archival Processing

The fourteen papers that detail how AI has been applied to archival processing can be divided into three areas: appraisal, arrangement and description, and preservation. Appraisal is the identification of materials for acquisition by an archive that have sufficient value. Arrangement and description are the organization and representation of records to achieve physical and intellectual control over the materials, and preservation refers to the protection of information from deterioration.

5.2.1. Appraisal

Five papers address the applications of AI for appraisal in archives. Colavizza et al. (2021), through their survey of the current uses of AI in archives, found that AI is automating appraisal decisions. Lappin (2020) also found that machine learning is being used to identify emails of significance. Fan et al. (2022) used machine learning to detect hate speech relating to COVID-19 in born-digital records for appraisal in a social media archive and proposed the use of Data-driven and Circulating Archival Processing as a way to process an evolving archive.

5.2.2. Arrangement and Description

Six papers discuss the applications of AI for arranging and describing archives. In addition to appraisal, Colavizza et al. (2021) also found that AI is automating the creation of metadata. Similarly, Bazán-Gil's (2023) analysis of case studies presented at audiovisual archives conferences from 2013 to 2023 found that AI is being used for the production of metadata and facial recognition in audiovisual archives. Meanwhile, Haffenden et al. (2023) found that two of the uses for NLP in the National Library of Sweden include the enrichment of metadata and the improved cohesion of digitized archival collections. Lombardi & Marianai's (2020) literature review also addressed the enrichment of metadata, detailing methods used to perform tasks like identifying creators and dating records. Randby & Marciano's (2020) case study from the Morgenthau Holocaust Collections Project demonstrated that machine learning could create a more usable finding aid. Their work aligns with Jaillant's (2022b), which proposes that using AI and the More Product Less Process approach for arrangement and description can increase access to email collection. Lastly, Davet et al. (2023) argue that by collecting, preserving, and evaluating paradata, a perceived gap in the "critical evaluation of archival arrangement and description" can be filled, and archival decisions may be redressed.

5.2.3. Preservation

Three papers discuss the use of AI for preservation. Kusumawati & Salim (2022) provide a literature review of the recent developments in using AI to preserve born-digital records, while the other two papers present case studies of effective applications. Modiba's (2022) research into the usability of AI for records management in the CSIR found that AI was able to maintain and preserve born-digital records. Regarding digitized records, Taurino & Smith (2022) present a case study from Northeastern University in which AI was used to digitize and recontextualize a collection of photographs to better preserve them.

## 5.3. Access and Use

Much of the literature, a total of thirty-two papers, focuses on using AI to increase the access and use of records. Subtheme one addresses the use of AI for privacy and detecting sensitive information in records. Subtheme two discusses the use of AI for enabling better record retrieval, and subtheme three explores how AI and records have been used to conduct research.

5.3.1. Privacy

Nine papers explored the subtheme of privacy. Three of the nine papers provide professional perspectives on this subtheme. Cushing & Osti's (2022) paper detailing the results of focus groups about the opinions of archivists on AI states that there is a lack of trust in AI systems to redact sensitive information accurately. Similarly, Li & Fleischmann's (2020) research, which explores the perspectives of ALA-accredited master's program students, found that many were concerned about the ethical challenges AI may pose to patron privacy. Conversely, Modiba's (2023) study on user perception of the use of AI for records management found that users think AI can facilitate adequate security of records. Three papers by Jaillant (2022a, 2022b) and one by Rees (2022), identify privacy restrictions as a major barrier to the access of born-digital and digitized records. In all three papers, using AI to identify sensitive information is proposed as a solution. Schneider et al. (2019) provide five case studies in which institutions are using the ML software ePADD to identify emails containing sensitive information. Two papers explore privacy as it relates to data collection. Jo & Gebru (2020) argue that ML for data collection could adopt methodologies from archives and libraries to address issues surrounding consent, power, inclusivity, transparency, ethics, and privacy, while Roberts & Montoya (2022) discuss privacy rights as they apply to the mining and sharing of data relating to Indigenous cultures.

5.3.2. Records Retrieval

The use of AI for searching and retrieving records was the most common application, with seventeen papers mentioning or discussing record retrieval. Six of the papers focus specifically on audiovisual records. Adewoye et al (2020). explore the potential of the Deep Archival Image Retrieval Engine (DAIRE), which retrieves images using visual content instead of text. Männistö et al. (2022) introduce Automatic Image Content Extraction (AICE), which can be used to analyze, classify, and search large image archives. Kumar et al. (2021) demonstrate how NLP techniques can be used to retrieve images from web sources. Bazán-Gil (2023) uses an analysis of case studies presented at audiovisual archives conferences from 2013 to 2023 and a survey to further explore the applications of AI in audiovisual archives, finding that it is often applied for retrieval, transcription, and translation purposes. Cheng et al. (2015) introduce an AI method that accurately recognizes and classifies scenes within videos in audiovisual archives, enabling better retrieval, and Yang (2023) explores the potential for AI to enhance text-to-video retrieval models.

Three papers address how AI can be used for the transcription of digitized records. Chrons & Sundell (2011) presented the crowdsourcing platform Digitalkoot, which used gamification to solve OCR errors during the transcription of digitized records, increasing their searchability and usability. Nockels et al. (2022) detail how the software Transkribus is being deployed, utilized, and reported in published research and establish a methodology for comparing handwritten text recognition platforms. Similarly, Terras (2022) explores how ML can be used to search, process, and generate transcriptions from mass-digitized manuscripts.

The last eight papers demonstrate how AI is used for retrieval in a variety of contexts. Suissa et al. (2021) explore the challenges of using deep neural networks to make records more available, searchable, and analyzable. Carter et al. (2022) used ML to improve the searchability of digitized diaries and papers in an under-utilized Holocaust collection, much like how Randby & Marciano (2020) used ML to create a more searchable finding aid as a part of the Morgenthau Holocaust Collections Project. Alothman & Wahab Sait (2022) provides a framework for using NLP to manage and retrieve Arabic and English bilingual documents. Colavizza et al. (2021), in a survey of AI in archives, also found that AI is used to index records based on content, enabling better retrieval. Ehrmann et al. (2023) provide a literature review of the approaches for Named Entity Recognition (NER) in digitized records, which identifies people, organizations, and locations of interest. Fan et al. (2022) used machine learning to retrieve records with hate speech relating to COVID-19 in born-digital records in a social media archive. Finally, Haffenden et al. (2023) found, during their exploration of the uses of NLP in the National Library of Sweden, that it can improve collections' searchability.

5.3.3. Use of Records in Research

Six papers provide examples of how records are being used in conjunction with AI to conduct research. Bottrighi et al. (2022) used machine learning and digital health records to predict whether or not patients had a high risk of mortality from COVID-19, Liu et al. (2023) used wind pressure records and machine learning to better understand the effects of wind pressure on high-rise buildings, and Michaud et al.(2023) used ML to reduce noise and improve classification in a bird song database. Jaillant (2022b), a literary scholar interested in accessing the email collections of authors, proposes using AI to increase access to such collections, while Suissa et al. (2021) provide a decision model for digital humanities researchers that details when and how to choose deep learning approaches for their research. Lastly, Okamoto et al. (2023) propose an approach for using digitized books and records to build accurate datasets for training machine learning models.

5.3.4. Public Access

In recent years, there has been a notable increase in the application of AI search technologies to manage public access requests under laws such as the UK and US Freedom of Information Acts. These AI solutions, commonly known as "technology assisted review" in e-discovery, are used to efficiently identify relevant documents and filter out sensitive information, including privacy-related and other privileged materials. Branting et al. (2023) developed an AI based method for automated detection of sensitive content in government records. Baron et al. (2022) applied AI techniques to identify Freedom of Information Acts-exempt material. These approaches address the critical challenge posed by the vast quantities of electronic records in modern archives, often referred to as "dark archives." Without AI, traditional keyword search methods struggle with accuracy, leading to excessive false positives and negatives and necessitating labor-intensive reviews. This situation raises a pressing question: What is the value of maintaining billions of records if they remain out of reach?

## 5.4. Professional Perspectives

The last theme, composed of five papers, contains literature about the opinions and perspectives of current and aspiring records managers, archivists, and information professionals on the implementation of AI in their field. Subtheme one specifically addresses AI in relation to archival education, while subtheme two explores the opinions of those in the field.

5.4.1. Archival Education

Three papers address how AI and other technologies are being taught in archival studies programs. Li & Fleischmann (2020) explored the perspectives of ALA-accredited master's program students on AI, and found that many are concerned about the potential ethical challenges AI may pose, such as patron privacy, censorship, algorithmic bias, and misinformation. Poole & Diaz (2022) investigated how North American archival studies programs are incorporating technology like AI, and found that there is a need to address these topics more in-depth. Similarly, Suissa et al. (2021) encouraged digital humanities and library and information studies departments to introduce mathematical and computer science topics like multivariable calculus, statistics, programming, deep learning, etc., into their academic syllabi.

5.4.2. User Perceptions

Two papers investigate how archivists, records managers, and other information professionals are thinking about AI. Cushing & Osti (2022) used focus groups to explore the opinions of archivists on AI and found that they were concerned about the potential additional workload and the potential to exacerbate bias against marginalized communities. Xie et al. (2022) studied how records and information management professional competencies align with the development of AI and found that while the profession is AI-resistant, it is not AI-proof.

# 6. Discussion

***RQ1*** - Section 4 of the paper addresses Research Question 1 (RQ1) by detailing how archivists and records management professionals have employed various AI techniques. The predominant methods are Natural Language Processing (NLP) and Computer Vision (CV), which are particularly suited for managing the extensive textual and visual data found in both paper archives and electronic records. Although many studies focus on automating processes through AI techniques, practicality remains a significant topic of debate. Our review found that 60% of the papers utilized pre-trained models for specific tasks. While these models are currently state-of-the-art, their memory footprint continues to hamper real-world applications. We believe that more work should be done to address inference issues rather than only concentrating on achieving high accuracy.

***RQ2*** - As detailed in Section 5, AI has been applied to most records management and archives functions. Within records management, the most common function being performed by AI is classification. Considering that classification is the organization of records into categories according to a plan or scheme (InterPARES Trust, 2023), this function can be easily carried out by AI, explaining the predominance of literature on the subject. It does not seem, however, that AI has been used to carry out functional analysis, the identification of the activities of an organization on which a classification schema is based. Within archives, AI has been fairly evenly applied to both appraisal and arrangement and description functions, with five and six papers on the subjects, respectively. Most of the literature, though, addresses the access and use of records, a function that occurs in both records management and archives. The use of AI for searching and retrieving records is by far the most prevalent use, with seventeen papers on the subject. Consequently, this suggests that, to date, the accessibility of records has been one of the primary problems addressed by AI. However, the focus on paper archives has been less frequent compared to digitized records (fourteen papers to twenty-two). This imbalance might be attributed to the predominance of digital records within records management environments.

***RQ3*** - As we have seen through the course of this study, many archives and records management functions can be performed by AI. By using AI to carry out menial, time-consuming tasks, archivists and other information professionals could use the excess time to address other creative and cognitively complex work, resulting in more records, both physical and digital, becoming available for access. There are also areas of archival work that AI has yet to be applied to or is just beginning to be applied to. For example, AI could detect outdated and offensive language in pre-existing catalogs that need to be changed or addressed in some way, or it could be used to flag quality control issues in the digitization process for audiovisual materials. Of course, generative AI will also provide information professionals with new challenges in how to best apply the technology, how to best acknowledge when and where it has been used, and how to manage AI-generated material in their repositories, particularly with the challenges it poses to traditional archival principles like creatorship. Regardless, AI has the potential to be an essential tool for dealing with the massive amount of digital records being created every day.

# 7. Conclusion and Future Directions

Not many studies have explored the role of AI in keeping digital records safe and readable over time. A key question is whether AI can help by automatically turning old or hard-to-read records into formats that are easy for everyone to use. While this might seem like a job for automation tools, AI could also play a crucial role. Although we have only just started to scratch the surface, this review has uncovered an interesting area to study: how people researching in archives use (and can use) AI to help with their work. Understanding this better could help us improve how archives operate in the future, making them more useful and accessible for everyone. One limitation of AI use is the ethical concerns surrounding the use of AI in archives, such as privacy issues and the potential for AI to introduce biases in how records are categorized. Additionally, the cost and technical expertise required to implement AI technologies can be significant barriers for many archival institutions, limiting their ability to adopt these methods.

In this systematic review of literature, we have analyzed 45 papers, written in English, and published between 2011 and 2023, on how AI can be used in archival science practices. One interesting observation through the course of this literature review was that the research in this area tends to split into two main types: the first focuses on how archives work, and the other dives deep theoretically into the AI side. The first category might not completely focus on how AI works, while the second is all about the AI model details without explaining how it fits into real-world archival work. This situation hints at a gap that could be bridged by more collaborations between archivists and computer scientists, suggesting a valuable opportunity for both fields to learn from each other and enhance how archives can benefit from advancements in AI.

The focus in AI applications for archives has been on Deep Learning and Natural Language Processing techniques. However, there is room to explore Reinforcement Learning, Contrastive Learning, and Large Language Models (LLMs) further. For instance, Reinforcement Learning could optimize archival retrieval systems by learning user search patterns to improve search results. Contrastive Learning might enhance the accuracy of categorizing and finding similarities between documents, making it easier to link related records. Lastly, LLMs could be employed to generate summaries of archival documents and answer user questions about documents and collections, therefore aiding researchers. These examples point toward untapped potential for enriching archival practices with advanced AI techniques.